\documentclass[a4paper,fleqn,usenatbib]{mnras}

\usepackage[T1]{fontenc}
\usepackage{ae,aecompl}

\usepackage[normalem]{ulem}
\usepackage{amsmath}
\usepackage{graphicx} 
\usepackage{lscape}
\usepackage{indentfirst}
\usepackage{enumitem}
\usepackage{xspace}

\usepackage{amssymb}
\usepackage[dvipsnames]{xcolor}
\usepackage{url}
\usepackage[flushleft]{threeparttable}

\newcommand {\bc}{\begin {center}}
\newcommand {\ec}{\end {center}}
\newcommand {\be}{\begin {equation}}
\newcommand {\ee}{\end {equation}}
\newcommand {\beq}{\begin {eqnarray}}
\newcommand {\eeq}{\end {eqnarray}}
\newcommand {\comment}[1]{}

\renewcommand{\d}{{\rm d}}

\newcommand {\ergs}{{\rm erg\ \rm s^{-1}}}

\title[Soft X-ray spectra in beamed XRPs]
{
Angle-dependent hardening of the reprocessed spectra in ULXs powered by accretion onto neutron stars
}
\author[S.~Karmakar, A.A.~Mushtukov \& M.~Middleton] 
{
Sricheta Karmakar,$^{1,2}$\thanks{Email: srichetak@gmail.com (SK)}
Alexander A. Mushtukov$^{3}$\thanks{Email: alexander.mushtukov@physics.ox.ac.uk (AAM)}
and
Matthew Middleton$^{4}$
\\ 
$^1$ School of Physics and Astronomy, University of Edinburgh, Edinburgh EH8 9YL, UK  \\
$^2$ Indira National School, Opp. Indira Institute, Off New Pune-Mumbai Highway, Wakad, Maharashtra 411057, India  \\
$^3$  Astrophysics, Department of Physics, University of Oxford, Denys Wilkinson Building, Keble Road, Oxford OX1 3RH, UK \\ 
$^4$ Department of Physics and Astronomy, University of Southampton, Highfield, Southampton SO17 1BJ, UK \\
}

\pubyear{2025}

\begin{document}
\label{firstpage}
\pagerange{\pageref{firstpage}--\pageref{lastpage}}

\maketitle


\begin{abstract} 
It is anticipated that mass accretion rates exceeding approximately $10^{19}\,{\rm g\,s^{-1}}$ in X-ray pulsars lead to radiation-driven outflows from super-critical accretion discs.
The outflows launched from the disc influence the angular distribution of X-ray radiation, resulting in geometrical beaming.
The beaming, in turn, impacts the apparent luminosity of the X-ray pulsar, detectability of pulsations, and the spectral composition of the X-ray flux.
We employ a straightforward geometrical model of the outflows, perform Monte Carlo simulations, and model the spectra of radiation, reprocessed by the walls of the accretion cavity formed by the outflows. We consider the \emph{reprocessed} emission only; direct pulsar emission is not included in our modelling.
Our results demonstrate that the spectra of reprocessed radiation depend on the actual luminosity of the central engine, the geometry of the outflows, and the viewing angle — most notably on the latter, through changing visibility of the hotter wall regions near the disc plane.
The high-energy part of the reprocessed spectrum depends strongly on viewing angle (harder at lower inclinations), while the soft flux varies comparatively little with inclination.
In our model, this contrast is a prediction: variable ultra-luminous X-ray sources are expected to exhibit strong high-energy angle sensitivity together with comparatively modest soft-band variation, naturally arising if precession modulates the effective inclination.
\end{abstract}


\section{Introduction}
\label{sec:Intro}

X-ray pulsars (XRPs) are accreting, strongly magnetised neutron stars in close binary systems (see \citealt{2022arXiv220414185M} for a review).
The magnetic field at the surface of a NS in an XRP is expected to be $\gtrsim 10^{11}\,{\rm G}$ \citep{2019A&A...622A..61S}.
In some XRPs, including the brightest XRP in our Galaxy Swift~J0243.6+6124 \citep{2022ApJ...933L...3K}, the NS magnetic field is reported to be $\gtrsim 10^{13},{\rm G}$ (see also \citealt{2020ApJ...899L..19G}).
The strong magnetic field shapes the geometry of the accretion flow within regions of size $\sim 10^8\,{\rm cm}$ around a NS, and channels the accretion flow towards small regions located close to the magnetic poles of the star.
Within the latter regions, accreting material is decelerated, kinetic energy turned into heat and predominantly emitted in the form of X-ray photons.
The apparent luminosity of XRPs spans a wide range, from $\sim 10^{32}\,\ergs$ up to $\sim 10^{41}\,\ergs$.
The brightest XRPs belong to the class of pulsating ultra-luminous X-ray sources (ULXs, see \citealt{2021AstBu..76....6F,2023NewAR..9601672K} for reviews), which are the focus of our research.
Intensive mass transfer in XRPs is expected to affect the geometry of the emitting regions, the structure of the accretion flow, and the formation of outflows:
\begin{itemize}[leftmargin=1em]
\item
At a mass accretion rate of $\dot{M} \gtrsim 10^{17}\,{\rm g\,s^{-1}}$, the radiative force becomes strong enough to halt the accretion flow above the NS surface, forming an accretion column — a structure supported by radiation pressure and confined by the strong magnetic field \citep{1976MNRAS.175..395B,1981A&A....93..255W,2015MNRAS.454.2539M,2023MNRAS.524.2431S,2023MNRAS.524.4148A,2025arXiv250602288Z}.
X-ray radiation from the accretion column experiences gravitational bending in the space-time curved by the NS, which affects (though not dramatically; see \citealt{2020PASJ...72...34I,2024MNRAS.527.5374M}) the apparent luminosity due to amplification of X-ray flux in certain directions.
At mass accretion rates of order $10^{19}\,\mathrm{g\,s^{-1}}$, accretion columns become advection-dominated, potentially producing strong, anisotropic neutrino emission \citep{2018MNRAS.476.2867M,2025MNRAS.538.2396M}.
\item At $\dot{M} \gtrsim 10^{18}\,{\rm g\,s^{-1}}$, the accretion flow within the magnetospheric radius becomes optically thick, influencing the processes of spectral and pulse formation \citep{2017MNRAS.467.1202M,2017A&A...608A..47K,2019MNRAS.484..687M,2023MNRAS.525.4176B,2024MNRAS.529.1571F}.
\item At $\dot{M} \gtrsim 10^{19}\,{\rm g\,s^{-1}}$, accretion discs in XRPs become radiation-pressure dominated and can lose mass through radiation-driven outflows \citep{1973A&A....24..337S,2007MNRAS.377.1187P,2019A&A...626A..18C}.
However, the stability of such radiation-driven outflows remains uncertain \citep{2022MNRAS.509.1119M}.
Numerical simulations of outflows in black hole binaries show that they can form accretion cavities with opening angles of $\phi \sim 0.5 - 1\,{\rm rad}$ \citep{2009MNRAS.398.1668O,2013PASJ...65...88T}.
In accreting, strongly magnetised NSs, most of the energy is released at the NS surface.
The outflow can collimate the X-ray flux, affecting both the apparent luminosity of XRPs \citep{2017MNRAS.468L..59K,2017MNRAS.466.2236D} and the pulse formation process \citep{2021MNRAS.501.2424M,2023MNRAS.518.5457M}.
In particular, strong geometrical collimation of X-ray flux in ULX pulsars is expected to suppress pulsations and cause phase shifts.
Evidence of powerful outflows has been reported in several bright XRPs and ULXs \citep{Pinto2016Natur.533...64P,2016ApJ...826L..26W,2018MNRAS.479.3978K,2019MNRAS.487.4355V}, with observed outflow velocities up to $\sim 0.2c$.
\end{itemize}
Thermally driven outflows, launched due to heating in the outer disc regions, may also arise in XRPs at comparatively low mass accretion rates (e.g., $\dot{M} \lesssim 10^{17}$ g s$^{-1}$, see \citealt{1983ApJ...271...70B,1996ApJ...461..767W,2015ApJ...807..107H}).
Such winds are typically initiated by Compton or photoionization heating and are expected to be important at larger radii where the escape velocity is lower.
These outflows may influence line formation and absorption features but are unlikely to significantly affect the inner accretion structure or beaming geometry.
Their effects are beyond the scope of this study but see \cite{2022MNRAS.509.1119M} for how thermal winds could disrupt accretion in super-Eddington accreting binaries.
Outflows launched in bright XRPs form an accretion cavity that collimates the X-ray flux in the direction close to the symmetry axis of the accretion flow (see Fig.\,\ref{pic:scheme}).
This collimation arises from multiple reflections and reprocessing of X-ray photons that interact with the cavity walls.
The dominant processes shaping this reflection are Compton scattering by electrons and free-free absorption/emission.
Due to absorption, a fraction of the X-ray flux illuminating the cavity walls is reprocessed — hard X-ray emission from the central object is absorbed and re-emitted in a softer energy band.
Such reprocessing by optically thick outflows is thought to produce the soft X-ray excess observed in ULX spectra \citep{2021PASJ...73..450K}.
Similar ideas were proposed in the context of ULXs hosting black holes, where reprocessing in a radiatively-driven wind funnel was shown to produce flat $\nu F_{\nu}$ spectra around $1$~keV and strong viewing-angle dependence \citep{2013AstBu..68..139V}. Although that study focused on supercritical accretion onto black holes, the general radiative effects of wind collimation and reprocessing are likely relevant for ULXs powered by strongly magnetised NSs as well.
Signs of X-ray photon reprocessing by outflows were recently reported in the Galactic X-ray binary Cyg~X-3 (see, e.g., \citealt{2024NatAs...8.1031V}). 
In what follows, we model the \emph{reprocessed} emission produced in the cavity walls; the \emph{direct} pulsar emission is not included in our calculations.
The observed X-ray emission in bright XRPs consists of both direct radiation from the central NS and radiation reprocessed by the outflows.
The relative contributions of these components depend on the outflow geometry and viewing angle of the distant observer. 
In this work we focus on the reprocessed component and its angle-dependent spectral properties.
Spectral variations correlated with the apparent luminosity have been reported in several ULXs, including ULX pulsars \citep{2014MNRAS.439.3461P,2017ApJ...836..113P,2015MNRAS.447.3243M, Walton2018ApJ...856..128W,2023A&A...672A.140F}.
Since the spectral shape is expected to depend on the viewing angle in systems with outflows (\citealt{2007MNRAS.377.1187P}), detection of such correlations supports the idea that outflows influence the X-ray emission of ULXs.
Some ULXs, including the brightest ULX pulsar NGC5907ULX1 \citep{2023A&A...672A.140F}, show super-orbital variability of X-ray flux on timescales of months, along with spectral changes correlated with the super-orbital phase.
These observations suggest that the super-orbital variability may be caused by precession of the accretion flow (see, e.g., \citealt{2002ApJ...565.1134S}) and the resulting changes in the observer’s viewing angle of the cavity formed by the outflows (e.g., \citealt{Middleton2018}, see also \citealt{2025arXiv250504523C}).
Although a qualitative model linking spectral shape to viewing angle has been proposed in the literature \citep{2015MNRAS.447.3243M,2017MNRAS.466.2236D,2018ApJ...853..115W}, we are not aware of any quantitative development of this idea.
In this paper, we consider a simple geometrical model of the outflows, and investigate the influence of collimation/geometrical beaming on the reprocessed-emission spectral formation. In particular, we quantify the viewing-angle–dependent spectral hardening of the reprocessed spectrum, while its soft flux varies comparatively weakly.
We also calculate the elements of the stress tensor at the boundaries of the accretion cavity, which will be essential for future self-consistent modelling of accretion cavity geometry in systems hosting strongly magnetised NSs.

\section{Model set up}
\label{sec:Model}

\begin{figure}
\centering 
\includegraphics[width=7.8cm]{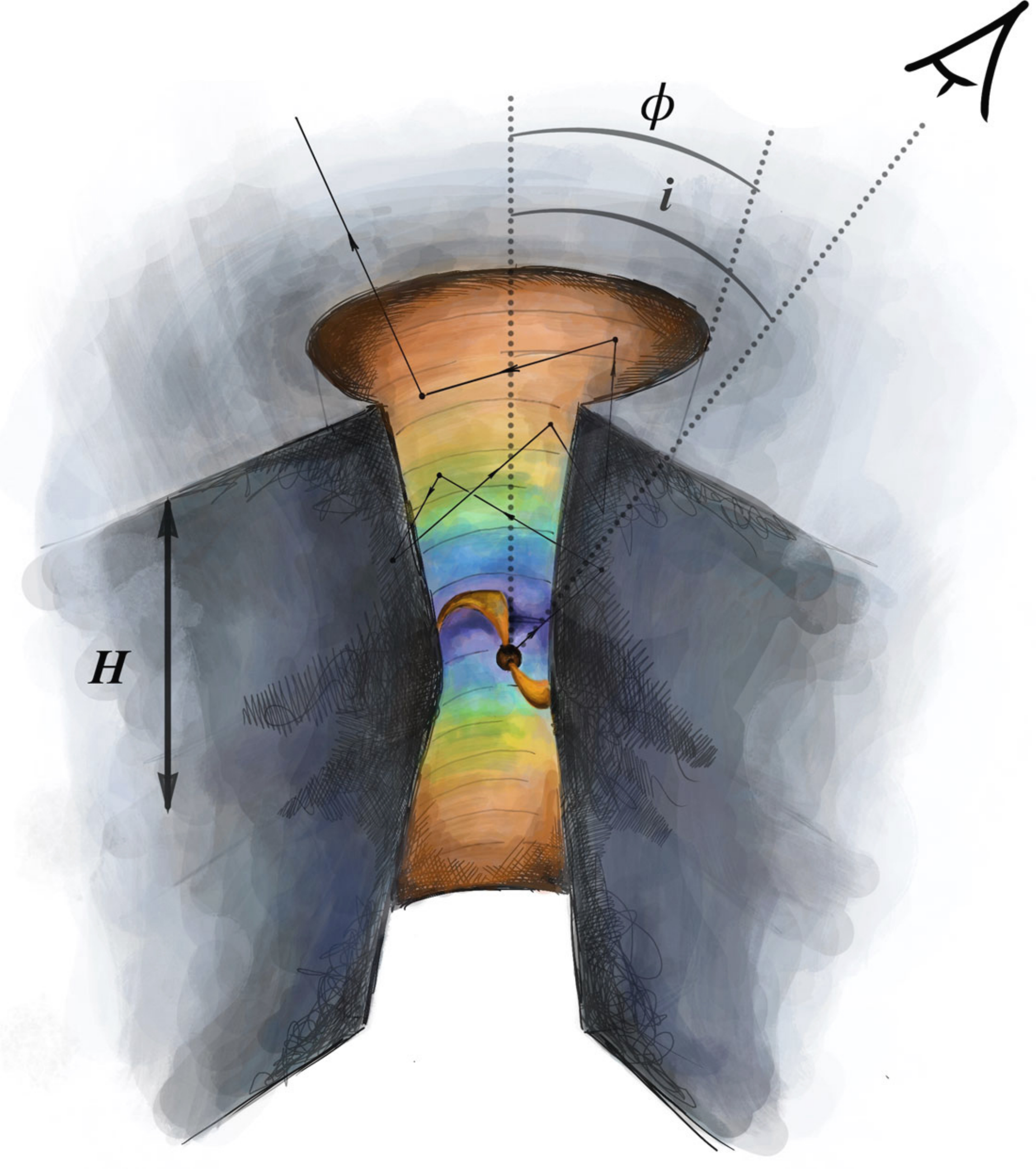} 
\caption{
A schematic picture of a bright XRP.
The central NS is located in an accretion cavity formed by the disc and the outflows. 
The geometry of the accretion cavity is determined by three parameters: the inner disc radius $R_{\rm m}$, the effective geometrical thickness of the cavity $H$, and its opening angle $\phi$.
The local temperature of the accretion cavity walls is determined by the amount of radiation that is reprocessed locally. 
The apparent luminosity and spectrum of the XRP is expected to be dependent on the viewing angle of a distant observer given by the inclination $i$.
}
\label{pic:scheme}
\end{figure}

We assume that the accretion flow in an XRP forms an accretion disc around a NS.
The disc is interrupted due to the interaction with the NS's magnetic field at the magnetospheric radius
\beq\label{eq:Rm}
R_{\rm m}=5\times 10^7\,\Lambda B_{12}^{4/7}\dot{M}_{19}^{-2/7}m^{-1/7}R_6^{12/7}\,\,{\rm cm},
\eeq
where $B_{12}$ is the magnetic field strength at the poles of a NS in units of $10^{12}\,{\rm G}$, $\dot{M}_{19}$ is the mass accretion rate from the disc in units of $10^{19}\,{\rm g\,s^{-1}}$, $m$ is a mass of the NS in units of solar masses, $R_6=R_{\rm NS}/10^{6}\,{\rm cm}$ is dimensionless NS radius, and $\Lambda$ is a dimensionless coefficient, which is typically taken to be $\Lambda=1$ for the case of accretion from the stellar wind and $\Lambda=0.5$ for the case of accretion from the disc (see \citealt{1978ApJ...223L..83G,1979ApJ...232..259G}, and estimated for the case of intensive mass transfer in \citealt{2019A&A...626A..18C}).
Unless stated otherwise, the mass accretion rate $\dot{M}$ refers to the accretion rate at the inner edge of the accretion disc, close to the magnetospheric radius.
Note that the magnetic field in (\ref{eq:Rm}) is assumed to be dipolar, although we treat later the inner disc radius as a free parameter to implicitly account for uncertainties in field strength and geometry.
The geometry of the outflows is given by the inner radius of the accretion flow (\ref{eq:Rm}), and two conical surfaces of opening angle $\phi$ and effective geometrical thickness $H$ (see Fig.\,\ref{pic:scheme}).
The material of the walls can move with dimensionless velocity $\beta=v/c$, which can affect the process of photon reprocessing from the walls of the accretion cavity.
We neglect rotation of the wind, assuming it to be subdominant compared to the outflow velocity. 
For the outflow speeds $\sim 0.1$--$0.2c$, rotational motion is not expected to significantly affect the dynamics or radiative transfer.
We assume that most of the energy is released close to the NS surface, which should be the case if $R_{\rm m}\gg R_{\rm NS}$.
Then X-ray photons start their travel in the accretion cavity from the central NS and leave the cavity straight away or after a few interactions with the cavity walls.
We neglect the influence of the flow between the inner disc radius and NS surface.

We perform Monte Carlo simulations, tracing the history of $N_{\rm tot}\sim 10^8$ photons and generating a series of random numbers $X_j\in(0;1)$ for each one.
Between interactions, photon packets propagate at fixed energy and frequency in the lab frame.
{The cavity interior is taken to be transparent; all interactions occur at the cavity wall. In this work we do not model Comptonization: the wall albedo is set to zero, and an incident packet is fully absorbed (its energy is deposited locally) and then thermally re-emitted. No optical-depth sampling inside the wall is performed.}
To estimate the physical conditions at the walls (i.e., local temperature and momentum exchange between the photons and accretion flow), we divide the wall into $2N_l$ layers of equal height 
$$\Delta h = \frac{H}{N_l}.$$
The simulation consists of the following steps: 
\begin{enumerate}[leftmargin=*]
\item \label{item:ph_emission}
The seed photons are emitted by a point source located at the origin of the coordinate system.
To get the direction of a seed photon, we generate random numbers $X_1,X_2$.
Then the initial direction of each photon is
\beq 
\theta_i = {\rm acos}(1 - X_1),\quad\quad 
\varphi_i = 2\pi X_2.
\eeq
\item\label{item:trajectory}
We assume that the photons are moving along straight trajectories within the accretion cavity and the photons either leave the system or reach some point at the conical surface.
Using the initial point of photon trajectory and direction of its motion, we check whether the photon crosses the walls of the accretion cavity or leaves the system.
If the photon crosses the cavity walls, we obtain the coordinates of the photon collision with the walls and go to step\,\ref{item:reflection}.
If the photon leaves the system, we account for its final direction of motion and return to step \ref{item:ph_emission} to consider the next photon. 
\item \label{item:reflection}
Upon reaching the conical surface, a packet is \emph{fully absorbed} and \emph{thermally re-emitted}.
In the local reference frame with the $z'$-axis orthogonal to the conical surface, the re-emitted photon direction is given by: 
\beq 
\theta' = {\rm asin}\sqrt{X_3},\quad\quad 
\varphi' = 2\pi X_4,
\eeq 
which corresponds to a Lambertian (cosine-law) angular distribution.
The unit vector describing the direction of the re-emitted photon in the local reference frame, where the $x$-axis is directed along the material velocity vector and $z$-axis is perpendicular to the accretion cavity surface, is Lorentz-transformed as
\beq 
\bf{n}'_\omega=  
\left(\begin{array}{c} 
(\sin\theta'\cos\varphi'-\beta)(1-\beta\sin\theta'\cos\varphi')^{-1} \\ 
D\sin\theta'\sin\varphi' \\
D\cos\theta'
\end {array}\right), 
\eeq
where
\beq\label{eq:D_gamma}
D&=&\gamma^{-1}(1-\beta\cos\chi)^{-1},\\
\gamma &=&(1-\beta^2)^{-1/2}
\eeq 
are the Doppler and gamma factors respectively and $\chi$ is the angle between the photon direction and the wall velocity in the comoving frame.
The unit vector of the photon direction in the laboratory reference frame is given by: 
\beq 
{\bf n}_\omega = \widehat{M}_z(\varphi_c - \pi)\widehat{M}_y\left(\frac{\pi}{2} - {\rm sgn}(z_c)\phi\right)\bf{n}'_\omega,
\eeq 
where $\widehat{M}_y(\alpha)$ and $\widehat{M}_z(\alpha)$ are rotation matrices.

{For energy accounting, each absorbed packet deposits its full energy in the local wall element (albedo $=0$); we keep a running tally of deposited energy and momentum. A new thermally re-emitted packet is launched in the comoving frame with the same packet energy, drawn from a Planck spectrum at the local temperature (set by energy balance; see below), and then Lorentz-transformed to the lab frame (frequency shift $\nu=D\,\nu'$).}

Simulating reprocessing of photons at the walls of the accretion cavity, we keep track of the number of collisions at each region of the accretion cavity and the momentum transferred to the walls due to the absorption–re-emission cycle:
\beq \label{eq:delta_p}
\Delta \mathbfit{p} = \mathbfit{p}_{\rm ini} - \mathbfit{p}_{\rm fin},
\eeq 
where $\mathbfit{p}_{\rm ini}$ and $\mathbfit{p}_{\rm fin}$ are momenta of a photon before and after interaction respectively.
The number of local interactions of photons with the wall and the amount of transferred momentum are used further to estimate the local temperature of the accretion cavity and the stress components respectively.
Then we return to step \ref{item:trajectory}.
\end{enumerate}

The local temperature of the walls of the accretion cavity is determined by the photon energy that is reprocessed locally. 
We consider heating of the outflowing material only by photons originating near the NS. 
Heating from the lower disc regions is neglected, which may lead to some underestimation of soft emission.
To get an approximate estimate, we assume that the temperature is related to the reprocessed flux as $F=\sigma_{\rm SB}T^4$.
We therefore obtain:
\beq \label{eq:temperature}
\sigma_{\rm SB}T_i^4 \simeq
L\frac{N_i}{N_{\rm tot}} \frac{1}{\pi \Delta h (R_{i+1}+R_i)},
\eeq 
where $L$ is the luminosity generated near the NS surface,
$N_i$ is the number of photon interactions (absorptions) within a layer of the accretion cavity, $N_{\rm tot}$ is the total number of photons participating in a simulation, $R_{i}$ and $R_{i+1}$ are the radii of the conical surface at the upper and lower borders of a layer, and $\Delta h$ is the geometrical thickness of a layer. 
Operationally, we iterate absorption and re-emission until both $T_i$ and the escaping spectrum change by less than $10^{-3}$ between successive iterations.

Multiple absorption–re-emission events of X-ray photons at the walls of the accretion cavity result in momentum transfer to the material of the outflow.
The momentum is transferred both in the direction perpendicular to the surface of the walls and along the walls, which causes stress normal to the wall's surface $\sigma_\perp$ and  stress parallel to the surface of the accretion cavity $\sigma_\parallel$.
The stress components are determined by the luminosity and the geometry of the accretion cavity and are dependent on the coordinates. 
Within the layer $i$, the stress components can be estimated as:
\beq\label{eq:stress}
\left(\begin{array}{c} 
\sigma_{i,\perp} \\ 
\sigma_{i,\parallel}
\end {array}\right) = 
\frac{1}{N_{\rm tot}S_i}
\frac{L_{\rm NS}}{c}
\sum\limits_{j}\Delta \mathbfit{p}_{j}\cdot
\left(\begin{array}{c} 
\mathbfit{n}_{i,\perp} \\ 
\mathbfit{n}_{i,b}
\end {array}\right),
\eeq 
where the summation is performed over all photons interacting with the layer, 
$\Delta \mathbfit{p}_{j}$ is the momentum transferred due to the interaction of a photon (\ref{eq:delta_p}),
$\mathbfit{n}_{i,\perp}$ is a local normal to the walls of the accretion cavity, and $\mathbfit{n}_{i,b}\perp \mathbfit{n}_{i,\perp}$ is the unit vector aligned with the velocity of the outflow.

As a result of our Monte Carlo simulations, we obtain maps of temperature and stress distribution over the walls of the accretion cavity.
The spectra detectable by a distant observer are composed of a direct component and the component reprocessed by the walls of the accretion cavity. 
The extracted temperature distribution gives us a chance to estimate the reflected part of the spectrum.
We assume that the wall emits locally blackbody spectra described by temperature $T_i$ estimated by (\ref{eq:temperature}):
\beq 
B_E(T)\simeq 5\times 10^{22}
\frac{E_{\rm keV}^3}{e^{E_{\rm keV}/kT}-1}\,\,
{\rm erg\,cm^{-2}\,s^{-1}ster^{-1}keV^{-1}}.
\eeq 
Different observers see different parts of accretion cavity and the spectrum of reprocessed radiation is given by a multi-color blackbody determined by the temperature distribution over the visible part: 
\beq 
F_E\propto \cos i\int\limits_{-R}^{R}\d x\int\limits_{-A}^{A}\d y\,B_E(T(x,y,i)),
\eeq 
where 
$R=R_{\rm in}+H\tan\phi$ and
$A=R\cos i \sqrt{1-(x/R)^2}$.

To clarify the scope of our model and facilitate interpretation of the results, we summarize below the main assumptions adopted throughout this work. These simplifications allow us to isolate the essential effects of geometrical beaming and radiative reprocessing in super-Eddington accretion flows:
\begin{itemize}[leftmargin=0.7em]
    \item The outflow is treated as steady, axisymmetric, and non-rotating. 
    Rotation is neglected under the assumption that the radial outflow velocity dominates over the azimuthal component in the cavity region.
    \item The cavity walls are fully ionized and optically thick to scattering and absorption. 
    We set the wall albedo to zero and assume full thermalization of incident X-ray packets at the wall (no Comptonization and no optical-depth sampling inside the wall). Re-emission is thermal (blackbody) from the local surface in the comoving frame and Lorentz-transformed to the lab frame.
    \item Reprocessing is modeled as perfect blackbody re-emission. 
    No color correction factor is applied; this may lead to an underestimate of the reprocessed temperature in the innermost regions.
    \item Special-relativistic bulk-motion effects (aberration and Doppler shift) are included; elastic scattering at the wall, thermal Comptonization, Klein–Nishina corrections, and gravitational redshift are neglected.
    \item Magnetic channelling between the inner disc and the NS surface is not included; instead, the model focuses on the region outside the magnetosphere, where outflows form and collimation occurs.
    \item No emission is included from regions outside the cavity.
\end{itemize}
These assumptions allow us to isolate and analyze the effects of cavity geometry and reprocessing on the observed spectra, while deferring more detailed microphysics to future work.

\section{Numerical results}
\label{sec:NumRes}
In this section, we first discuss the geometrical collimation of X-ray flux (see Section,\ref{sec:Res_collimation}), then we present the calculated distributions of radiation pressure over the walls of the accretion cavity (Section,\ref{sec:Res_pressure}), temperature distribution (Section,\ref{sec:temperature}) and, finally, we discuss the expected shape of spectra formed due to multiple reprocessing of X-ray photons (Section,\ref{sec:Res_sp}). 
For orientation: geometrical beaming trends are shown in Fig.\,\ref{pic:beam01} (panels a–c), with spectra discussed later in Section\,\ref{sec:Res_sp}.

\subsection{X-ray collimation}
\label{sec:Res_collimation}

\begin{figure}
\centering
\includegraphics[width=7.7cm]{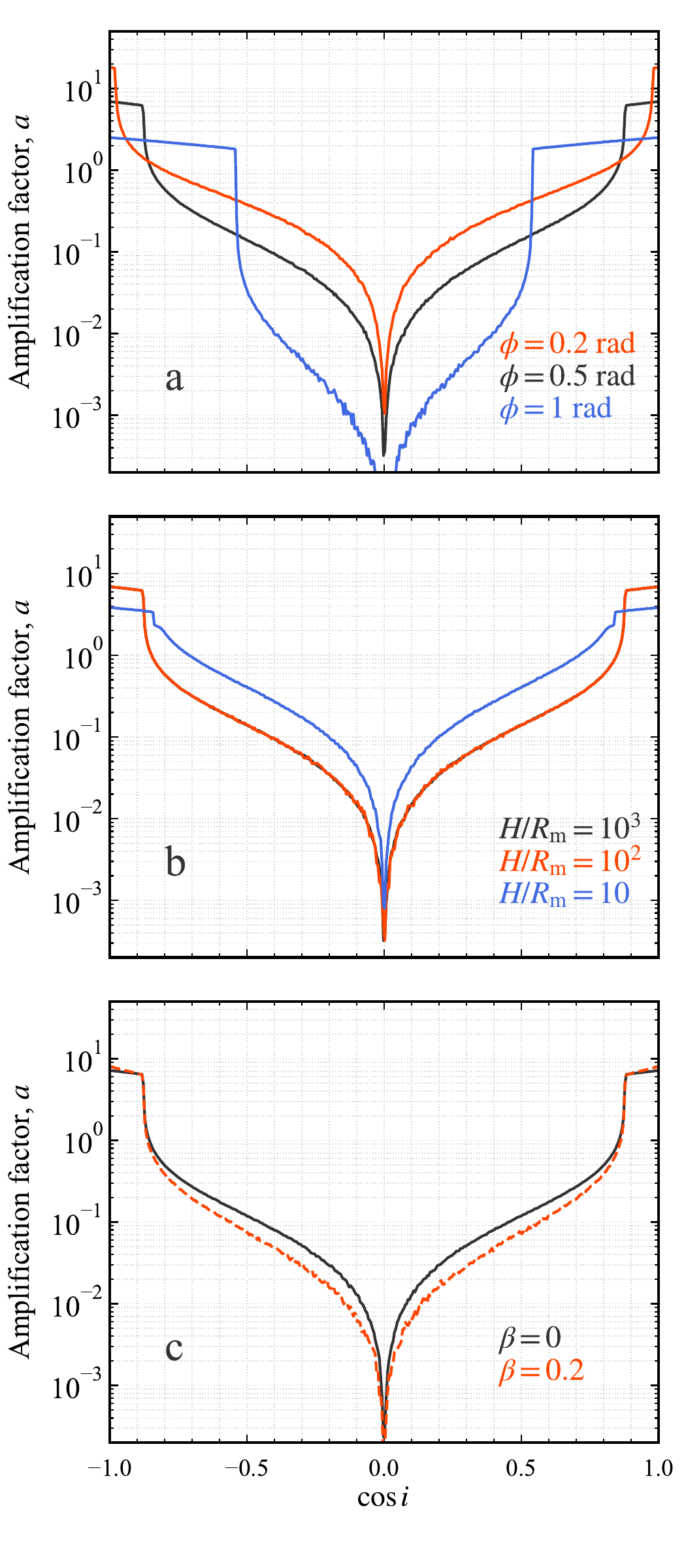}
\caption{
Angle-dependent geometrical beaming of the emission (see Section\,\ref{sec:Res_collimation}).
The amplification factor as a function of observer inclination $i$ with respect to the accretion cavity.
(a) The accretion cavity relative height and wall velocity are fixed at $H/R_{\rm m}=10^3$ and $\beta=0$. Different curves correspond to various opening angles of the cavity:
{$1\,{\rm rad}$} (blue line),
{$0.5\,{\rm rad}$} (black line),
{$0.2\,{\rm rad}$} (red line).
(b) The opening angle and wall velocity are fixed at {$\phi=0.5\,{\rm rad}$} and $\beta=0$. Different curves correspond to varying relative heights of the accretion cavity:
$H/R_{\rm m}=10^3$ (black line),
$H/R_{\rm m}=10^2$ (red line),
$H/R_{\rm m}=10$ (blue line).
(c) The relative height and opening angle are fixed at $H/R_{\rm m}=10^3$ and {$\phi=0.5\,{\rm rad}$}. Different curves correspond to different velocities of the accretion cavity walls:
$\beta=0$ (black solid line),
$\beta=0.2$ (red dashed line).
}
\label{pic:beam01}
\end{figure}

Beaming of X-rays in the considered geometry is dependent both on the opening angle of the accretion cavity $\phi$ and its effective geometrical thickness $H$ (see Fig.\,\ref{pic:beam01}a–b).
The smaller the opening angle and the larger the ratio $H/R_{\rm m}$, the larger the amplification factor within the opening angle of the cavity (cf.\ Fig.\,\ref{pic:beam01}a for varying $\phi$ and Fig.\,\ref{pic:beam01}b for varying $H/R_{\rm m}$).
Note that a significant fraction of photons leave the system with $i>\phi$ when a distant observer detects only the flux reflected/reprocessed by the walls {(see the $i>\phi$ regime in Fig.\,\ref{pic:beam01}a–b)}.
This reduces the maximal amplification factor and the apparent luminosity:
\beq
L_{\rm app}
<
\left(\frac{2\pi}{\Omega}\right)L
\eeq
even for an observer looking at the system along the axis of symmetry {(examples in Fig.\,\ref{pic:beam01}a–b; see also numerical simulations in} \citealt{2017MNRAS.466.2236D}).
Within the cavity opening angle, the apparent luminosity always exceeds the isotropic luminosity and the fraction of the direct flux is related to the amplification factor as:
\beq
f_{\rm dir}=a^{-1},
\eeq
while, out of the opening angle, the apparent luminosity is typically smaller than the isotropic one and $f_{\rm dir}=0$ (compare the $i<\phi$ and $i>\phi$ branches in Fig.\,\ref{pic:beam01}a–b).
According to our simulations, the maximal amplification factor for the case of an accretion cavity with opening angle,
$0.1,{\rm rad}<\phi<1,{\rm rad}$,
and relative height of the cavity,
$20<H/R_{\rm m}<10^3$
can be roughly approximated as:
\beq\label{eq:approx}
a_{\rm max}\approx 6.8\,e^{-(\phi/1\,{\rm rad})}
\left(\frac{\phi}{1\,{\rm rad}}\right)^{0.1[\log_2(0.1H/R)+1]}.
\eeq
{This approximation has been validated against our grid of models shown in Fig.\,\ref{pic:beam01}a–b (see the peak values of $a$ as a function of $\phi$ and $H/R_{\rm m}$).}
The effective motion of the walls of the accretion cavity (i.e., $\beta>0$) makes X-ray reflection anisotropic, which reduces the flux and amplification factor for observers with $i>\phi$ and increases the amplification for observers with $i<\phi$ (see Fig.\,\ref{pic:beam01}c).
These results are in agreement with earlier numerical analyses performed by \citealt{2017MNRAS.466.2236D} and \citealt{2023MNRAS.518.5457M}.

\subsection{Radiative pressure on the walls of the accretion cavity}
\label{sec:Res_pressure}

\begin{figure}
\centering
\includegraphics[width=8.7cm]{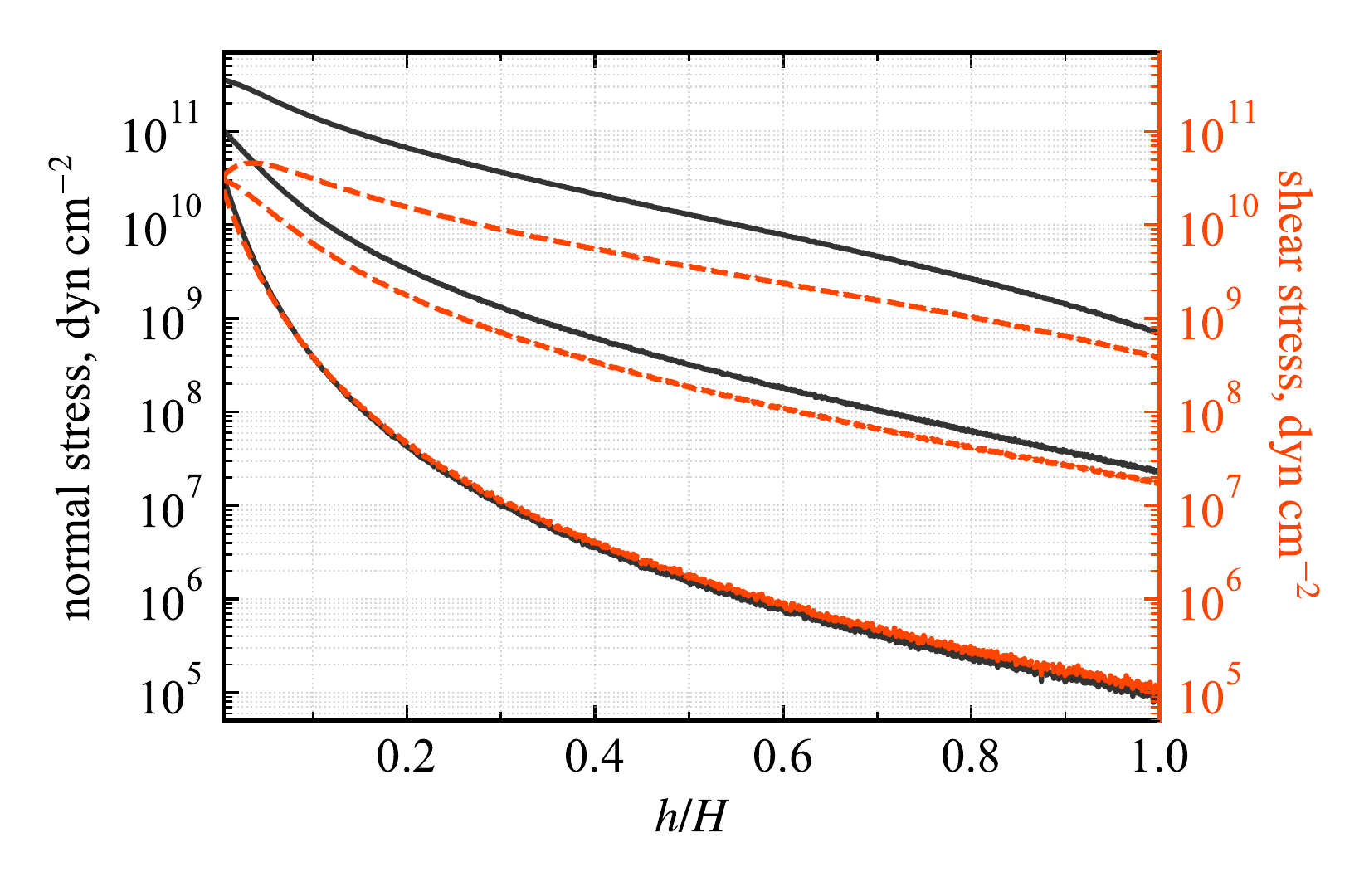}
\caption{
Radiation-induced stress on the cavity walls (see Section\,\ref{sec:Res_pressure}).
The normal (black solid lines) and shear (red dotted lines) stress at the walls of the accretion cavity.
Different curves are given for different opening angles of the cavity (from bottom to top):
{$\phi=1\,{\rm rad}$,}
{$0.5\,{\rm rad}$,}
{$0.2\,{\rm rad}$.}
The smaller the opening angle, the larger the stress at the walls.
Parameters:
$L=5\times 10^{39}\,\ergs$, 
$R_{\rm m}=5\times 10^7\,{\rm cm}$, 
$H/R_{\rm m}=10$.
}
\label{pic:sigma_1}
\end{figure}

\begin{figure}
\centering
\includegraphics[width=8.7cm]{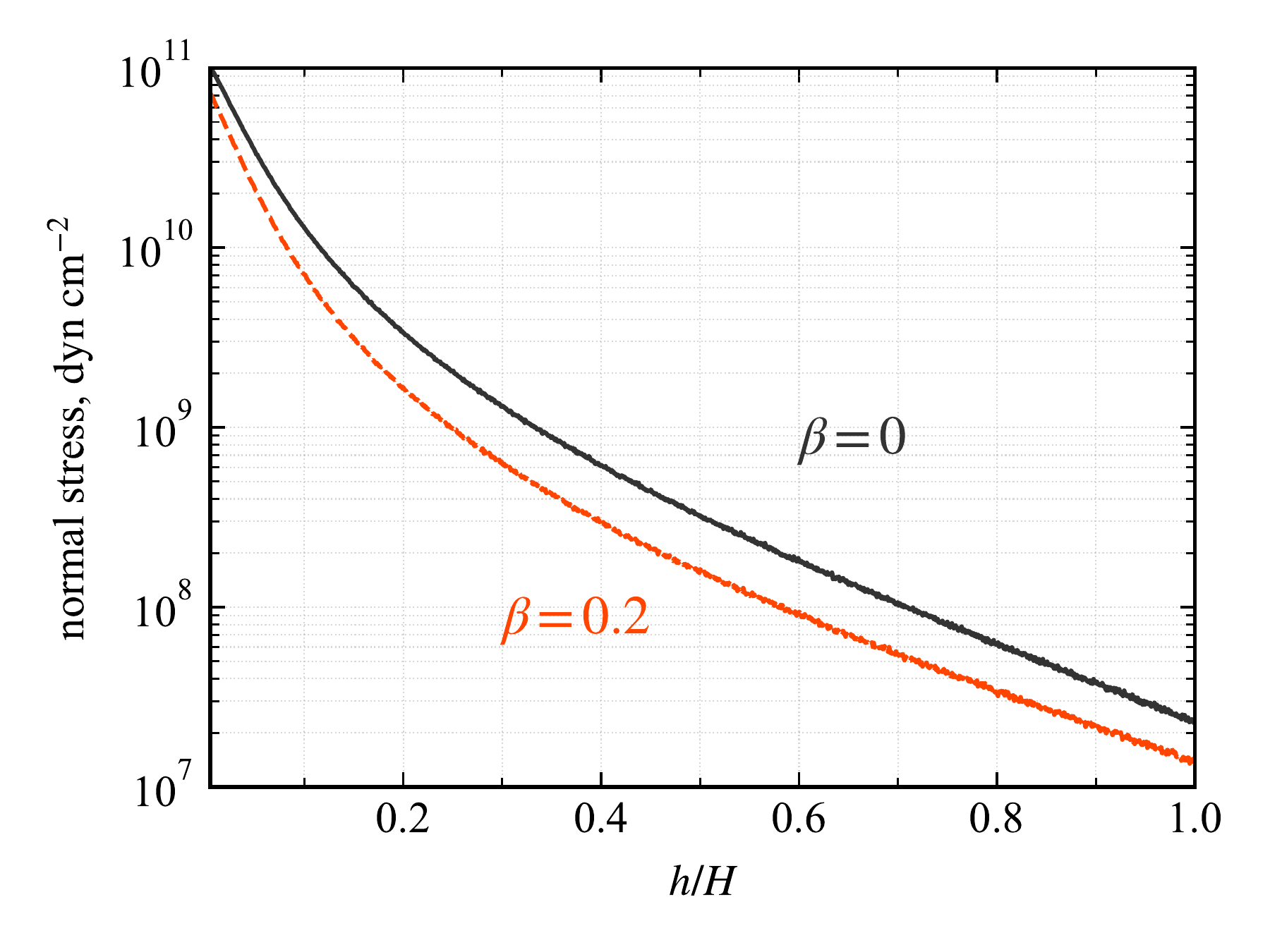}
\caption{
Effect of wall motion on the normal stress (cf.\ Fig.\,\ref{pic:sigma_1}).
Normal stress at the walls of the accretion cavity calculated for different wall velocity: $\beta=0$ (black solid line) and $\beta=0.2$ (red dashed line).
One can see that including effects of special relativity reduces the stress by a factor $\sim 2$ for the case of $\beta=0.2$.
Parameters:
$L=5\times 10^{39}\,\ergs$, $R_{\rm m}=5\times 10^7\,{\rm cm}$, $H/R_{\rm m}=10$, $\phi=0.5\,{\rm rad}$.
}
\label{pic:sigma_2}
\end{figure}

Photon oscillations between the walls of the accretion cavity result in momentum transfer that can influence the geometry (see Section 3.4 in \citealt{2023MNRAS.518.5457M}).
For a given and fixed geometry of the accretion cavity, we calculate the normal $\sigma_\perp$ and shear $\sigma_\parallel$ stress as a function of height $h$ (see expression \ref{eq:stress}; the resulting profiles are shown in Fig.\,\ref{pic:sigma_1} and Fig.\,\ref{pic:sigma_2}).
For opening angles $\phi\lesssim 1\,{\rm rad}$, the normal stress typically exceeds the shear stress across most of the wall height (see Fig.\,\ref{pic:sigma_1}).
Both normal and shear stresses tend to be larger for smaller opening angles {(compare the $\phi=1\,{\rm rad}$, $0.5\,{\rm rad}$, and $0.2\,{\rm rad}$ curves in Fig.\,\ref{pic:sigma_1})}.
Accounting for the effects of special relativity due to motion of material at the walls of the accretion cavity tends to reduce the stress for a given geometry; this reduction is illustrated in Fig.\,\ref{pic:sigma_2}.

\subsection{Temperature of the walls of the accretion cavity}
\label{sec:temperature}

\begin{figure}
\centering 
\includegraphics[width=8.7cm]{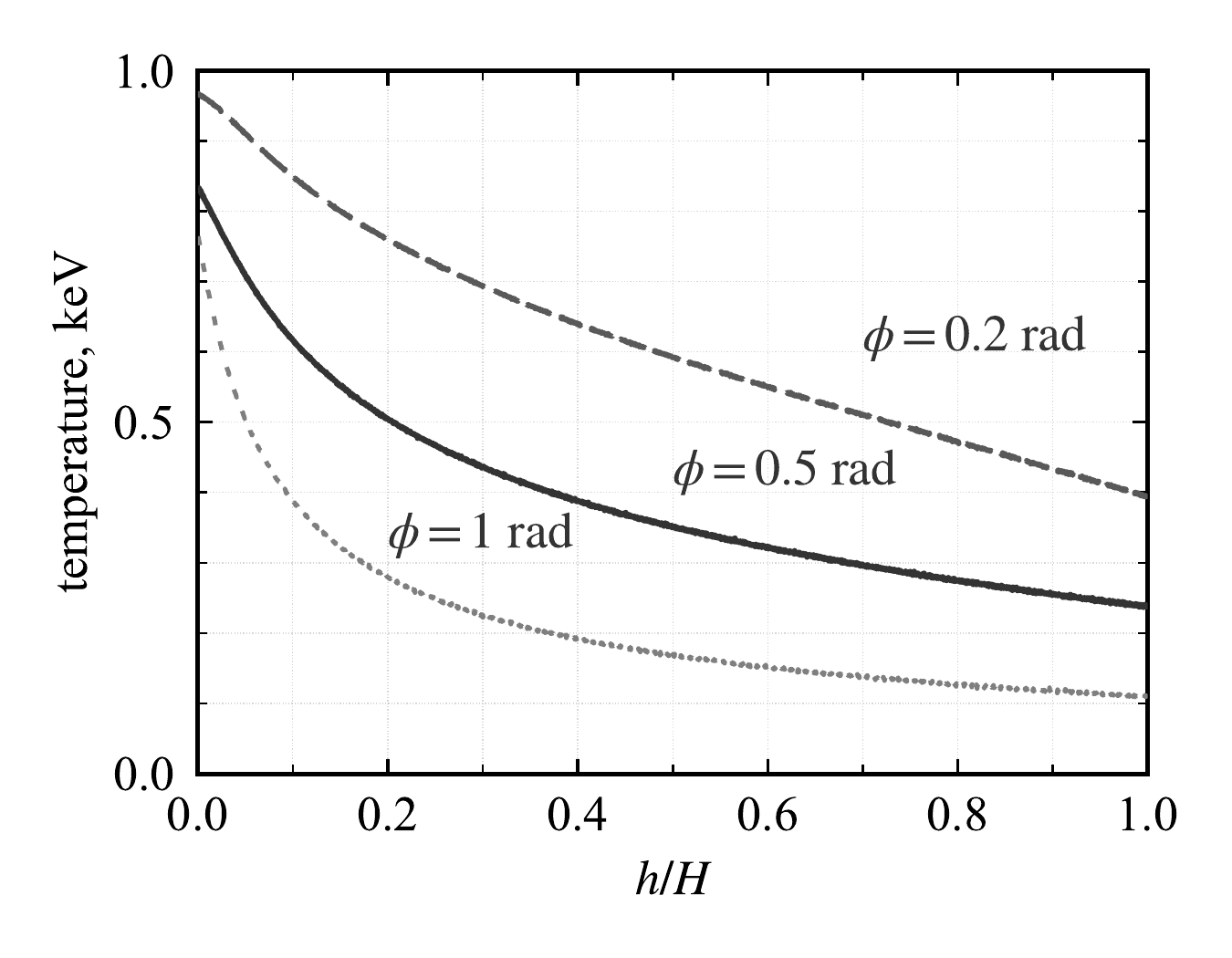} 
\caption{
{Temperature profiles over the cavity walls for different opening angles (see Section\,\ref{sec:temperature}).}
The temperature distribution over the walls of the accretion cavity.
Different curves are shown for different opening angles of the cavity: 
{$\phi=0.2\,{\rm rad}$,}
{$0.5\,{\rm rad}$,}
{and $1\,{\rm rad}$ (from top to bottom).}
The temperature of the walls tends to increase towards the accretion disc plane.
The smaller the opening angle, the higher the temperature of the walls.
Parameters: 
$L=5\times 10^{39}\,\ergs$, $R_{\rm m}=5\times 10^7\,{\rm cm}$, $H/R_{\rm m}=10$.
}
\label{pic:temp01}
\end{figure}

\begin{figure}
\centering 
\includegraphics[width=8.5cm]{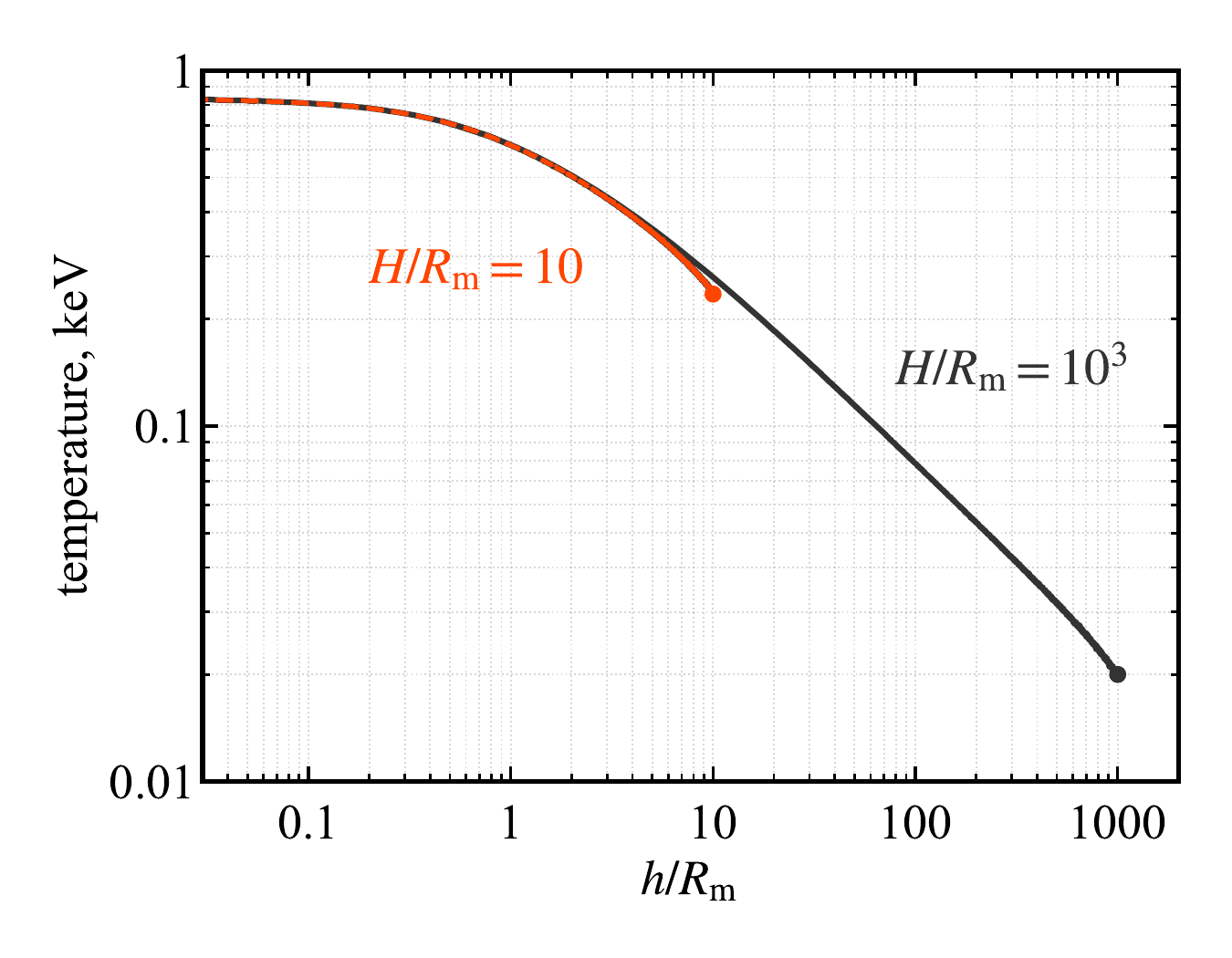} 
\caption{
Effect of cavity thickness on temperature profiles (cf.\ Fig.\,\ref{pic:temp01}).
The temperature distribution over the walls of the accretion cavity.
Red and black curves are given for different relative heights of the accretion cavity: $H/R_{\rm m}=10$ and $H/R_{\rm m}=10^3$ respectively.
Parameters: 
$L=5\times 10^{39}\,\ergs$, $R_{\rm m}=5\times 10^7\,{\rm cm}$, $\phi=0.5\,{\rm rad}$.
}
\label{pic:temp02}
\end{figure}

The temperature distribution over the surface of the accretion cavity is related to the amount of energy locally reprocessed by the walls.
For a given geometry of accretion cavity (i.e., fixed $\phi$, $R_{\rm m}$, $H/R_{\rm m}$ and $\beta$) the temperature scales with the luminosity of the central source as $T\propto L^{1/4}$, consistent with the trends shown in Fig.\,\ref{pic:temp01}–\ref{pic:temp02}.
The effective temperature tends to decrease with distance from the accretion disc plane (see Fig.\,\ref{pic:temp01}).
The larger the opening angle of the cavity, the steeper the temperature decrease.
Because the number of local reflections increases with a decrease in accretion cavity opening angle, the temperature tends to be higher in the case of smaller $\phi$  (see Fig.\,\ref{pic:temp01}).
The larger geometrical thickness of the accretion cavity results in the appearance of regions of lower temperature (see Fig.\,\ref{pic:temp02}), but temperature behavior in regions located close to the accretion disc plane is similar in the case of the same opening angle $\phi$ (compare red and black curves in Fig.\,\ref{pic:temp02}). 
In the case of $H/R_{\rm m}\gg 1$ and $h\lesssim R_{\rm m}$, the temperature is weakly dependent on the coordinate:
\beq 
T\approx 0.64\,\phi^{-0.15}
\left(\frac{L}{10^{40}\,\ergs}\right)^{1/4}
\left(\frac{R_{\rm m}}{10^8\,{\rm cm}}\right)^{-1/2}
\,{\rm keV}.
\eeq 
At $h>R_{\rm m}$ the temperature dependence on $h$ can be approximated by a power law: 
\beq
T&\approx& 
1.7\,e^{-1.9\phi}
\left(\frac{h}{R_{\rm m}}\right)^{-0.63\phi^{0.15}}\\
&&\nonumber
\times\left(\frac{L}{10^{40}\,\ergs}\right)^{1/4}
\left(\frac{R_{\rm m}}{10^8\,{\rm cm}}\right)^{-1/2}\,{\rm keV},
\eeq 
i.e., at larger opening angles of the accretion cavity, the temperature at the walls drops faster with $h$; the accuracy and trends of these approximations are illustrated in Fig.\,\ref{pic:temp01} (varying $\phi$) and Fig.\,\ref{pic:temp02} (varying $H/R_{\rm m}$).

\subsection{Spectra of reprocessed radiation}
\label{sec:Res_sp}

\begin{figure}
\centering 
\includegraphics[width=8.7cm]{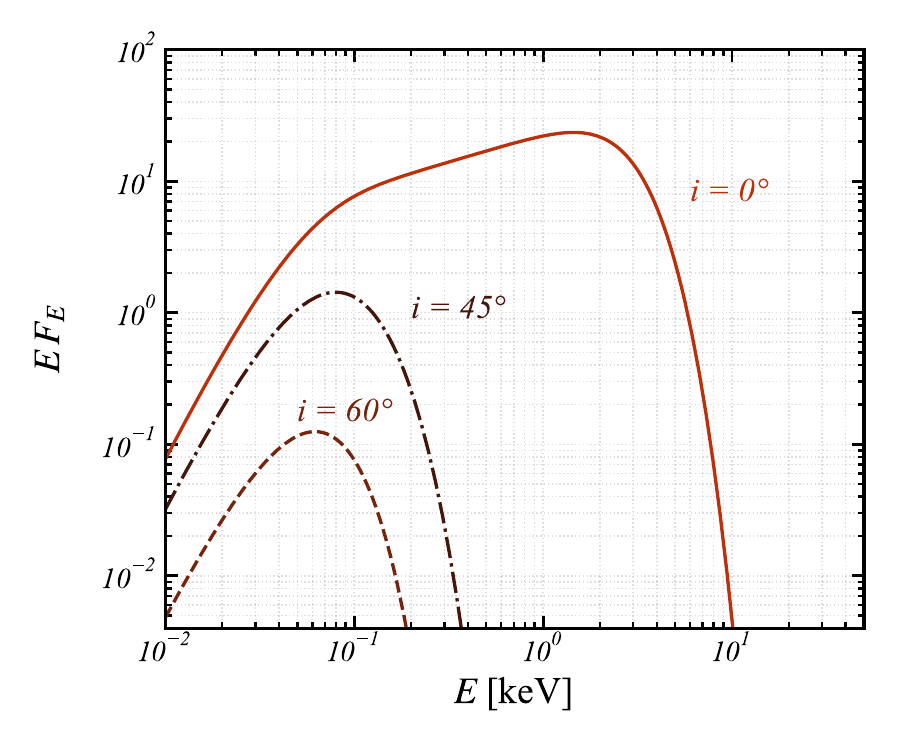} 
\caption{
{Angle dependence of the \emph{reprocessed} spectrum from the accretion cavity.
Model parameters: $L=5\times10^{39}\,\mathrm{erg\,s^{-1}}$, $R_{\rm m}=10^8\,\mathrm{cm}$, $H/R_{\rm m}=10^3$, $\phi=0.5\,\mathrm{rad}$, and $\beta=0$.
Curves show observer inclinations $i=0^\circ$ (solid), $45^\circ$ (dash–dot), and $60^\circ$ (dashed), i.e. a “low vs.\ high inclination” view.
The high-energy part hardens strongly toward lower $i$ (greater visibility of hotter wall regions near the disc plane), while the soft flux varies comparatively little with inclination.
Only the reprocessed component is shown; direct pulsar emission is not included.}
}
\label{pic:spec_00}
\end{figure}

\begin{figure*}
\centering 
\includegraphics[width=15.5cm]{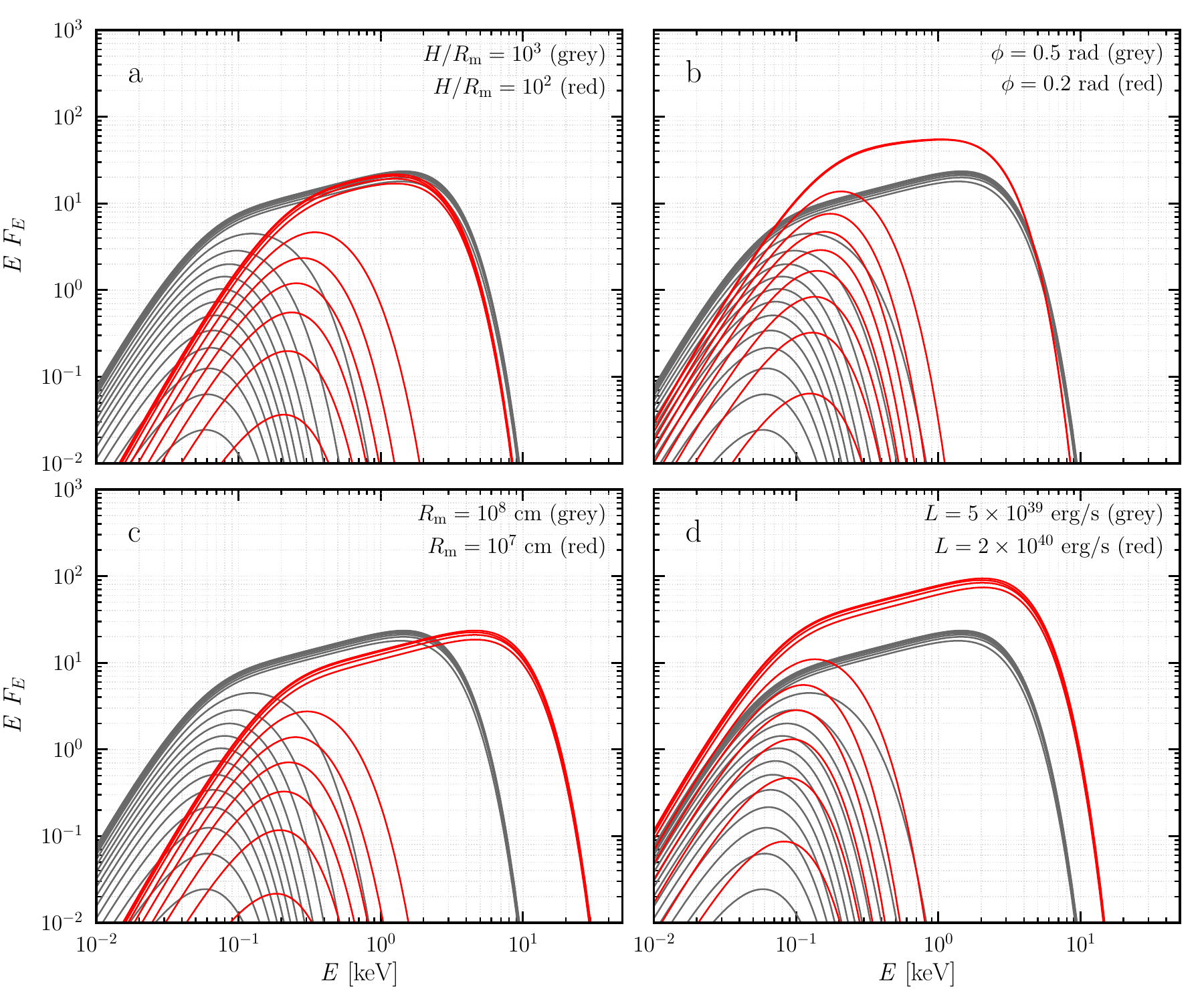} 
\caption{
Spectra of radiation, reprocessed by the accretion cavity for different model parameters. 
Black curves in all panels correspond to the fiducial model with 
    $L = 5\times10^{39}$\,erg\,s$^{-1}$, 
    $R_{\rm m} = 10^8$\,cm, 
    $H/R_{\rm m} = 10^3$, 
    $\phi = 0.5$\,rad, and 
    $\beta = 0$. 
Each black line represents a different observer inclination angle with 
$\cos i = 0.95,\,0.85,\,0.75,\,0.65,\,0.55,\,0.45,\,0.35,\,0.25,\,0.15,\,0.05$ (top to bottom).
\textit{Panel (a):} Red curves show spectra for $H/R_{\rm m} = 10^2$. 
\textit{Panel (b):} Red curves show spectra for $\phi = 0.2$\,rad. 
\textit{Panel (c):} Red curves show spectra for $R_{\rm m} = 10^7$\,cm. 
\textit{Panel (d):} Red curves show spectra for $L = 2\times10^{40}$\,erg\,s$^{-1}$.
Across each panel, lower-inclination curves show enhanced high-energy flux (harder reprocessed spectra), while the soft flux is comparatively stable with inclination.
}
\label{pic:spec}
\end{figure*}

The observed X-ray emission from bright XRPs includes the direct component from the central source and the radiation reprocessed by the walls of the accretion cavity. Observers situated within the cavity's opening angle (i.e.\ $i<\phi$) detect both components, whereas observers outside this angle (i.e.\ $i>\phi$) see only the reprocessed emission. Our modelling focuses solely on the reprocessed component, which is governed by the geometry of the cavity and the visibility of its surfaces from different inclinations (see Fig.\,\ref{pic:spec} for parameter trends and Fig.\,\ref{pic:spec_00} for a compact low–vs–high inclination comparison).

The resulting reprocessed spectra resemble a multi-colour blackbody and exhibit two characteristic breaks. The low-energy break corresponds to the minimum temperature attained at the upper parts of the cavity, while the high-energy break reflects the maximum temperature near the disc plane, where irradiation is strongest; both break energies shift with geometry and luminosity as expected from the wall-temperature scalings in Section\,\ref{sec:temperature}.

The spectral hardness depends on the viewing angle: lower inclinations (i.e. closer to the symmetry axis) provide greater visibility of the hotter regions near the disc plane, producing harder spectra, whereas the soft reprocessed flux varies comparatively weakly with inclination. Thus, a correlation between spectral hardness and apparent luminosity naturally arises, which could be observable as periodic or quasi-periodic spectral variability driven by precession of the cavity or outflow (illustrated by the inclination sequence in Fig.\,\ref{pic:spec}).

The spectra are further modulated by the cavity geometry. {Panel (b) of Fig.\,\ref{pic:spec}} shows the effect of decreasing the cavity opening angle from $\phi=0.5$ to $0.2$\,rad, and panel (a) shows the effect of reducing the relative cavity height from $H/R_{\rm m}=10^3$ to $H/R_{\rm m}=10^2$. In both cases, the reprocessed spectra become softer overall due to changes in the temperature structure along the cavity wall {(cf.\ the trends in Fig.\,\ref{pic:temp01}–\ref{pic:temp02})}.

The bottom panels in Fig.\,\ref{pic:spec} illustrate the impact of the inner disc radius and accretion luminosity. Panel (c) compares models with $R_{\rm m}=10^8$ and $10^7$\,cm, showing that smaller magnetospheres lead to hotter cavity walls and hence harder spectra. Panel (d) demonstrates the temperature shift due to increased X-ray luminosity from $L=5\times10^{39}$ to $2\times10^{40}$\,erg\,s$^{-1}$, which moves both spectral breaks to higher energies due to the local scaling ($T \propto L^{1/4}$) {(see also Section\,\ref{sec:temperature})}.

Overall, the viewing angle, cavity geometry, and luminosity each influence the spectral shape of reprocessed emission in distinct and predictable ways. Across all cases, lower inclinations enhance the high-energy part of the \emph{reprocessed} spectrum (spectral hardening), while the soft flux remains comparatively stable with inclination, offering a route to interpret inclination-dependent spectral variability in ULX pulsars.

\section{Summary}

We have considered the process of geometrical beaming and spectral formation in bright X-ray pulsars, where the mass accretion rate is high enough to produce radiation-driven outflows from the accretion disc \citep{1973A&A....24..337S,2019MNRAS.484..687M,2019A&A...626A..18C}.
In such systems, a significant fraction of photons emitted in close proximity to the NS are reprocessed and collimated by the outflows. This process alters both the apparent luminosity and the spectral shape observed from different viewing angles. 
Throughout this work we model the \emph{reprocessed} emission produced in the cavity walls; the \emph{direct} pulsar emission is not included.
We extended earlier work on geometrical beaming in bright X-ray pulsars \citep{2021MNRAS.501.2424M,2023MNRAS.518.5457M} by developing a Monte Carlo model that follows photon trajectories, their interactions with cavity walls, and momentum transfer within the reprocessed component.
An important analytical result of our work is the approximate expression for the maximum luminosity amplification factor due to beaming (\ref{eq:approx}), which captures how the observed luminosity depends on the cavity geometry and the viewing angle.
This complements our numerical simulations, which demonstrate how reprocessing shapes the emergent reprocessed spectra for different observer inclinations.
Assuming that X-ray photons are mostly absorbed and re-emitted by the cavity walls, we estimated the temperature distribution over the wall surface (Section~\ref{sec:temperature}).
The temperature peaks near the disc plane and decreases with height, resulting in multi-color blackbody spectra with energy-dependent angular variation (see Fig.\,\ref{pic:temp01} and \ref{pic:temp02}).
The model predicts that spectra become systematically softer at larger inclinations - i.e., when the observer's line of sight lies farther from the cavity axis — consistent with spectral variability observed in ULXs (e.g., \citealt{2015MNRAS.447.3243M, 2020MNRAS.491.5702P}); equivalently, the reprocessed spectra harden toward lower inclinations, while the soft reprocessed flux varies comparatively weakly with viewing angle.
This supports the idea that spectral changes can be caused by variations in the viewing angle, for instance due to precession of the accretion disc \citep{2015MNRAS.447.3243M, 2017MNRAS.466.2236D}.
We also computed the local stress tensor components (normal and tangential) on the walls of the accretion cavity (Section,\ref{sec:Res_pressure}). These quantities are essential for constructing a self-consistent picture of cavity dynamics and geometry, potentially enabling future coupling with radiation hydrodynamics or magnetohydrodynamics.
Our findings connect well with observations of sources like NGC 55 ULX \citep{2017MNRAS.468.2865P}, which show transitions between ultraluminous X-ray and supersoft states likely caused by changing wind optical depth and inclination \citep{2015MNRAS.447.3243M}.
The observed softening at higher inclination, and the presence of residuals near 1~keV in their spectra \citep{Middleton2014MNRAS.438L..51M,Middleton2015MNRAS.454.3134M, Pinto2016Natur.533...64P, Kosec2021MNRAS.508.3569K}, align with our model predictions of reprocessing-dominated spectra in high-inclination systems.
The relative enhancement of the high-energy flux within the reprocessed component at low inclinations is also naturally explained in our framework, as higher-energy photons experience fewer interactions and can escape more freely along the axis; we emphasise that any additional hard emission from the direct pulsar component lies outside the scope of our modelling.
We note, however, that our model does not include emission from the extended, quasi-spherical wind outside the cavity.
At large inclinations, this additional component may dominate the observed flux, producing very soft spectra typical of so-called super-soft ULXs \citep[e.g.,][]{2007MNRAS.377.1187P,2015MNRAS.454.3134M,2016MNRAS.456.1837S,2016MNRAS.456.1859U}. Including this contribution would be important for modeling the full range of ULX spectral types and will be considered in future work.
Our model predicts that a decrease in the apparent luminosity of a ULX can naturally be accompanied by a softening of its X-ray spectrum — a direct consequence of increasing viewing angle through a collimating cavity.
At low energies, the emergent spectrum follows $\nu F_\nu \propto \nu^3$, consistent with a Rayleigh--Jeans-like slope commonly observed in ULX spectra.
For example, NGC5907ULX1 - a confirmed NS ULX - exhibits a steep low-energy rise in $\nu F_\nu$ below 2~keV \citep{2017ApJ...834...77F}, in qualitative agreement with our predictions.
Although we do not perform detailed spectral fitting, our results capture key observational trends, lending further support to the role of reprocessing and geometrical beaming in shaping ULX spectra.
In certain configurations, our model also produces nearly flat spectra in the soft X-ray band, with $\nu F_\nu \propto \nu^0$ (see red curves in Fig.\,\ref{pic:spec}b).
This behavior results from the angular distribution of temperature on the cavity walls and resembles spectra expected from supercritical accretion models such as SCAD \citep{2013AstBu..68..139V}.
The ability to reproduce such flat spectra - also seen in ULX data - further supports the relevance of our reprocessing-based geometrical model.
Although our current model neglects magnetic channeling between the magnetosphere and the NS surface, and assumes simplified reprocessing physics (e.g., full thermalization), it provides a robust starting point for interpreting the spectra of ULX pulsars and other highly beamed systems; moreover, the direct pulsar emission is intentionally excluded from the present treatment.
Future improvements could include partial albedo treatment and direct comparison with time-resolved spectra and timing data of transitional sources.
In summary, we confirm that observed spectral variability in ULXs - including softening at lower luminosity phases - can be explained by changes in the observer’s viewing angle relative to a geometrically beamed, reprocessing-dominated system.
Specifically, the viewing-angle dependence manifests primarily as \emph{spectral hardening} of the \emph{reprocessed} emission toward lower inclinations, while the soft reprocessed flux remains comparatively stable with angle, naturally linking precession-driven inclination changes to the observed spectral states.

\section*{Acknowledgements}

The authors thank Alexandra Veledina and Juri Poutanen for motivating discussions.
AAM thanks UKRI Stephen Hawking fellowship. 
We are grateful to an anonymous referee for their useful comments and suggestions.

\section*{Data availability}

The calculations presented in this paper were performed using a private code developed and owned by the corresponding author. All the data appearing in the figures are available upon request. 


\bsp 
\label{lastpage}
\end{document}